\documentclass[a4paper,12pt]{article}
\usepackage[cp1251]{inputenc}
\usepackage{amsmath}
\usepackage{amsthm}
\usepackage{amssymb}
\usepackage{graphics}
\usepackage{cite}
\usepackage[matrix,arrow]{xy}
\usepackage{trush_vol}

\theoremstyle{definition}
\newtheorem{definition}{Definition}
\newtheorem*{example}{Example}
\theoremstyle{theorem}
\newtheorem{theorem}{Theorem}
\DeclareMathOperator{\Tr}{Tr} 

\begin{document}

\title{\Large\textbf{General model of quantum key
distribution}}

\author{\normalsize{A. S.
Trushechkin$^{1,2}$ and I. V.
Volovich$^1$}}\date{\normalsize\textit{$^1$Steklov Mathematical
Institute of the Russian Academy of Sciences, Moscow\\
$^2$Moscow Engineering Physics Institute, Russia\\
e-mail: trushechkin@mail.ru, volovich@mi.ras.ru}} \maketitle

\begin{abstract}
A general mathematical framework for quantum key distribution
based on the concepts of quantum channel and Turing machine is
suggested. The security for its special case is proved. The
assumption is that the adversary can perform only individual (in
essence, classical) attacks. For this case an advantage of
quantum key distribution over classical one is shown.
\end{abstract}

\section{Introduction}

In a number of papers several concrete quantum key distribution
protocols were suggested, most of them are based on the BB84
protocol \cite{BennettBrassard84}. Its security is considered,
e.g., in
\cite{Mayers,LoChau,Biham,ShorPreskill,Gisin,VolovichVolovich}.
Practical realizations of quantum key distribution protocols are
described, e.g., in \cite{Gisin,Zeilinger}. However, formal
mathematically rigorous general approach to quantum key
distribution is lacking at present. For recent discussions of
these problems see \cite{BenorMayers,RennerEkert,RennerGisin}.

In this paper we suggest a rather general mathematical model of
quantum key distribution. We also prove the security for its
certain special case. The security is proved on the assumption
that the adversary can perform only individual (in essence,
classical) attacks. For this case an advantage of the quantum key
distribution over classical one is shown.

\section{Notations}

$\mathcal H$ is a Hilbert space, $\mathcal{S(H)}$ is a convex set
of quantum states (density operators) on $\mathcal H$. Let
$\mathcal H_A$ and $\mathcal H_B$ be a pair of Hilbert spaces. A
channel $\Theta$ from $\mathcal H_A$ to $\mathcal H_B$ is an
affine map from $\mathcal{S(H}_A)$ to $\mathcal{S(H}_B)$ such that
its linear extension has a completely positive conjugate map
\cite{OhyaPetz,Holevo}. The sequence of channels
$$\Theta^n:\mathcal{S(H}_A^{\otimes
n})\to\mathcal{S(H}_B^{\otimes n}),~~n\in\mathbb N$$ is associated
with the channel $\Theta$ by the formula
$$\Theta^n(\rho_1\otimes\ldots\otimes\rho_n)=\Theta(\rho_1)\otimes
\ldots\otimes\Theta(\rho_n),
~~\rho_i\in\mathcal{S(H}_A),~~i=1,\ldots,n.$$ If $\mathcal A$ and
$\mathcal B$ are finite sets, $\mathcal{P(A)}$ and
$\mathcal{P(B)}$ are probability distributions on $\mathcal A$ and
$\mathcal B$, then an affine map
$V:\mathcal{P(A)}\to\mathcal{P(B)}$ specifies a classical channel.
If $P\in\mathcal{P(A)}$, then $I(P,V)$ denotes the Shannon mutual
information between input and output of the channel $V$ if $P$ is
a distribution on input of the channel $V$. A channel $V$ can be
also specified by a mapping of the corresponding random values. If
$X$ and $Y$ are random values, then $I(X;Y)$ denotes the mutual
information between them. By
$V^n:\mathcal{P(A}^n)\to\mathcal{P(B}^n)$ denote the discrete
memoryless channel corresponding to the channel $V$. An affine map
$\Xi:\mathcal{P(A)}\to\mathcal{S(H)}$ can be specified by a
function $\xi:\mathcal A\to\mathcal{S(H)}$. Let
$\rho\in\mathcal{S(H)}$ be a quantum state. Then the von Neumann
entropy is defined by the formula $$H(\rho)=-\Tr\rho\log\rho.$$
Let $\mathcal A$ be a finite set, $P\in\mathcal{P(A)}$ a
distribution, $\mathcal H$ a Hilbert space, $\xi:\mathcal
A\to\mathcal{S(H)}$ a function. We define
$$C(\xi)=\max_{P}\left[H\left(\sum_{a\in\mathcal
A}P(a)\xi(a)\right)-\sum_{a\in\mathcal A}P(a)H(\xi(a))\right].$$
If $\{M(b)\}$ is a positive operator-valued measure (POVM), i.e.,
an observable, on $\mathcal H$, then the formula $$P(b)=\Tr
M(b)\rho,~~\rho\in\mathcal{S(H)}$$ specifies an affine map (a
channel) from $\mathcal{S(H)}$ to $\mathcal{P(B)}$. The space of
POVM on $\mathcal H$ taking values on $\mathcal B$ denotes by
$\mathcal{M(H;B)}$. An observable from $$\mathcal{M(H};\mathcal
B)^{\otimes n}\subset\mathcal{M(H}^{\otimes n};\mathcal B^n)$$ we
will call a factorized observable on the space $\mathcal
H^{\otimes n}$ taking values on $\mathcal B^n$. An observable that
doesn't belong to this class we will call an entangled observable.
By $$\mathcal{B^A}\circ\mathcal{M(H;A)}\subset\mathcal{M(H;B)}$$
denote the class of observables of the form
$$\{F(b)=\sum_{a\in f^{-1}(b)}M(a)\}_{b\in\mathcal
B},~~\{M(a)\}_{a\in\mathcal A}\in\mathcal{M(H;A)},$$ where $f$ is
an element of the set $\mathcal{B^A}$ of functions from $\mathcal
B$ to $\mathcal A$.

\section{General model of quantum key distribution}

In this section a definition of general mathematical model of
quantum key distribution is outlined. We consider the following
problem of key distribution. Two parties, Alice and Bob, want to
get a pair of keys (one key for Alice and another one for Bob)
using communication channels. A realization of a certain random
value on a finite set $\mathcal K$, or this random value itself is
regarded as a key. If Alice's and Bob's keys are identical with
high probability, and Eve's information about the keys is
negligibly small, then the problem of key distribution is
considered to be solved with some \textit{security degree}.

We will model the parties using extended Turing machines, which
form the model of classical computers interacting with external
devices and communicating with each other by classical and quantum
channels.

Let $(\mathcal A,\mathcal Q_A,\tau_A)$ be a Turing machine
describing Alice, where $\mathcal A$ is an alphabet, $\mathcal
Q_A$ is a set of states with additional chosen elements, and
$\tau_A$ is a transition function. Further, let $(\mathcal
A,\mathcal Q_B,\tau_B)$ and $(\mathcal A,\mathcal Q_E,\tau_E)$ be
Turing machines describing Bob and Eve accordingly.

\begin{definition}
A \textit{system of quantum key distribution} is a triple of
objects (\textit{extended Turing machines})

$$(\mathrm{ETM}_A,\mathrm{ETM}_B,\mathrm{ETM}_E),$$
where
$$\mathrm{ETM}_A=(\mathcal A,\mathcal Q_A,\tau_A,P_A,\xi_A),$$
$$\mathrm{ETM}_B=(\mathcal
A,\mathcal Q_B,\tau_B,P_B,\{\xi_B^{(n)}\}_{n\in\mathbb N}),$$
$$\mathrm{ETM}_E=(\mathcal
A,\mathcal Q_E,\tau_E,P_E,\{\xi_E^{(n)}\}_{n\in\mathbb
N},\mu_E).$$ Here $P_A,P_B,P_E$ are probability distributions on
$\mathcal A^+=\bigcup_{i=1}^\infty\mathcal A$, $$\xi_A:\mathcal
A^+\to\mathcal
S_A\subset\bigcup_{i=1}^\infty\mathcal{S(H}_A^{\otimes i}),$$
$$\xi_B^{(n)}:\mathcal A^+\to\mathcal
M_B^{(n)}\subset\mathcal{M(H}_B^{\otimes n};\mathcal A^+),$$
$$\xi_E^{(n)}:\mathcal A^+\to\Omega^{(n)},$$
$$\mu_E:\mathcal
A^+\to\mathcal M_E\subset\mathcal{M(H}_E;\mathcal A^+),$$ where
$\mathcal H_A,\mathcal H_B,\mathcal H_E$ are Hilbert spaces,
$\Omega^{(n)}=\{\Theta^n_i\}_{i\in I}$,
$$\Theta^n_i:\mathcal{S(H}_A^{\otimes n}\otimes\mathcal
H_E)\to\mathcal{S(H}_B^{\otimes n}\otimes\mathcal H_E).$$ Upon the
functions $\xi_A,\xi_B^{(n)},\xi_E^{(n)},\mu_E$ are imposed some
restrictions. They will be presented in the further paper.
\end{definition}

Interaction of the parties by quantum and classical channels is
realized by the use of the additional chosen elements in $\mathcal
Q_A$, $\mathcal Q_B$, $\mathcal Q_E$ and the functions introduced
above.

\section{Special model of quantum key distribution}

In this paper we will consider only a special case of the
described model.

\begin{definition}

A \textit{system} $G$ \textit{of quantum key distribution} is a
family of the following objects:

\begin{equation}\label{EqScheme}G=
\left(\mathcal K,\mathcal H_A,\mathcal H_B,\mathcal
H_E,\Theta,\{q^{(n)}\}_{n\in\mathbb N},\{M_B^{(n)}\}_{n\in\mathbb
N},\{\mathcal M_E^{(n)}\}_{n\in\mathbb N}\right).\end{equation}
Here $\mathcal K$ is a finite set (a set of keys), $\mathcal
H_A,\mathcal H_B,\mathcal H_E$ are Hilbert spaces,
$$\Theta:\mathcal{S(H}_A)\to\mathcal S(\mathcal H_B\otimes\mathcal
H_E)$$is a channel. The functions $$q^{(n)}:\mathcal
K\to\mathcal{S(H}_A^{\otimes n})$$ specify channels $Q^{(n)}$,
$$M_B^{(n)}\in\mathcal M(\mathcal H_B^{\otimes n};\mathcal K),~~
\mathcal M_E^{(n)}\subset\mathcal M(\mathcal H_E^{\otimes
n};\mathcal K).$$

\end{definition}

For any $n\in\mathbb N$ and $M_E^{(n)}\in\mathcal M_E^{(n)}$ we
define the channel $$\Lambda_n=(M_B^{(n)}\otimes
M_E^{(n)})\circ\Theta^n\circ Q^{(n)}$$ with the input alphabet
$\mathcal K$ and the output alphabet $\mathcal K^2$.

Let $K_A$ denote a random value uniformly distributed on $\mathcal
K$ (a key). We denote $K_B$ and $K_E$ random variables taking
values on $\mathcal K$ and related to $K_A$ by the channel
$\Lambda_n$ for some $n$, i.e., $$(K_B,K_E)=\Lambda_n(K_A).$$ The
schematic view of the quantum key distribution process is
presented on Fig. 1.
\newpage

$$
\xymatrix{K_A\ar[r]&\mathcal
H_A\ar[r]\ar[dr]&\mathcal H_B\ar[r]&K_B\\
&&\mathcal H_E\ar[r]&K_E}
$$
\begin{center}{Fig. 1}\end{center}

The proposed model is a quantum analogue of the classical key
distribution model considered in \cite{CsiszarKorner78}, where the
following theorem is proved.

\begin{theorem}\label{TheoCsiszar}Let $\mathcal{K,A,B,E}$ be finite sets,
and a pair of channels
$$V:\mathcal{P(A)}\to\mathcal{P(B)},~~W:\mathcal{P(A)}\to\mathcal{P(E)}$$
is given. Further, let
\begin{equation}\label{EqCCCond}\max_{P\in\mathcal{P(A)}}[I(P,V)-I(P,W)]>0.\end{equation}
Then for any $\alpha,\beta\in(0,1)$ and any sufficiently large
$n\in\mathbb N$ there exists a channel (a random coder)
$$F_A:\mathcal{P(K)}\to\mathcal{P(A}^n)$$ and a function (a decoder)
$$f_B:\mathcal B^n\to\mathcal K$$ such that for any function (a
decoder) $$f_E:\mathcal E^n\to\mathcal K$$ we have:

1) $\Pr[K_A=K_B]\geq\alpha,$

2) $I(K_A;K_E)\leq 1-\beta,$\\
where $$K_B=f_B\circ V^n\circ F_A(K_A),~~K_E=f_E\circ W^n\circ
F_A(K_A).$$

If for any $P$ the condition $$I(P,V)\geq I(P,W)$$ is true, then
condition (\ref{EqCCCond}) is not only sufficient, but also
necessary for existing of $F_A$ and $f_B$ with the specified
properties for sufficiently large $n$.

\end{theorem}

So, condition (\ref{EqCCCond}) can be considered as a condition of
possibility of classical key distribution with an arbitrary
security degree. A pair of numbers $(\alpha,\beta)$ is regarded as
a security degree.

Now we formulate our main theorem.

\begin{theorem}\label{TheoOur}
Let us fix $\mathcal H_A,\mathcal H_B,\mathcal H_E,\Theta$ in the
model $G$ of quantum key distribution (\ref{EqScheme}). For any
$n\in\mathbb N$ we put $$\mathcal M_E^{(n)}= \mathcal K^\mathcal
E\circ\mathcal M(\mathcal H_E;\mathcal E)^{\otimes n},$$ where
$\mathcal E$ is a finite set.

Suppose there exist a finite set $\mathcal A$ and a channel $\Xi$
specified by a function $\xi:\mathcal A\to\mathcal S(\mathcal
H_A)$ that obey the following property:

\begin{equation}\label{EqQCCond}C(\Theta_B\circ\xi)>C_1(\Theta_E\circ\xi),
\end{equation}
where $\Theta_B=\Tr_{\mathcal H_E}\Theta$, $\Theta_E=\Tr_{\mathcal
H_B}\Theta$, $$C_1(\Theta_E\circ\xi)=\max_{P\in\mathcal{P(A)},
M\in\mathcal{M(H}_B;\mathcal K)}I(P,M\circ\Theta_E\circ\Xi).$$
Then for any $\alpha,\beta\in(0,1)$ and any sufficiently large
$n\in\mathbb N$ there exist a channel (a random coder)
$$F_A:\mathcal{P(K)}\to\mathcal{P(A}^n)$$ and an observable
$$M_B\in\mathcal M(\mathcal H_B^{\otimes n};\mathcal K)$$ such that
$\forall M_E^{(n)}\in\mathcal M_E^{(n)}$ the random values $K_A$,
$K_B$ and $K_E$, where $(K_B,K_E)=\Lambda_n(K_A)$, obey the
properties:

1) $\Pr[K_A=K_B]\geq\alpha$,

2) $I(K_A;K_E)\leq 1-\beta$.\\
Here $$\Lambda_n(K_A)=(M_B^{(n)}\otimes
M_E^{(n)})\circ\Theta^n\circ Q^{(n)},$$ $$Q^{(n)}=\Xi^n\circ
F_A.$$
\end{theorem}

\begin{proof}[Outline of proof]Let us denote
$$C_k(\Theta_B^k\circ\xi^{(k)})=\max_{P\in\mathcal{P(A)},
M\in\mathcal{M(H}^{\otimes k}_B;\mathcal K)}
I(P,M\circ\Theta_B^k\circ\Xi^k),$$ where $$\xi^{(k)}:\mathcal
A^k\to\mathcal{S(H}_A)$$ is the function that specifies the
memoryless channel $\Xi^k$ corresponding to the channel $\Xi$,
i.e.,
$$\xi^{(k)}(a_1,\ldots,a_k)=\xi(a_1)\otimes\ldots\otimes\xi(a_k).$$
As \cite{Holevo}
$$C(\Theta_B\circ\xi)=\lim_{k\to\infty}\frac{1}{k}C_k(\Theta_B^k\circ
\xi^{(k)}),$$ then, in view of (\ref{EqQCCond}), $$\exists n~
C_n(\Theta_B^n\circ \xi^{(n)})>nC_1(\Theta_E\circ\xi).$$ Since for
any $k$, in view of the conditions on the adversary's measurements
class,
$$\max\limits_{\stackrel{P\in\mathcal{P(A}^k)}
{M\in\mathcal{M(H}_E;\mathcal E)^{\otimes k}}}
I(P,M\circ\Theta_E^k\circ
\xi^{(k)})
=kC_1(\Theta_B\circ\xi),$$ this implies that
$$\max\limits_{\stackrel{P\in\mathcal{P(A}^n)}
{M\in\mathcal{M(H}^{\otimes n}_B;\mathcal K)}}
I(P,M\circ\Theta_B^n\circ
\xi^{(n)})
>
\max\limits_{\stackrel{P\in\mathcal{P(A}^n)}
{M\in\mathcal{M(H}_E;\mathcal E)^{\otimes n}}}
I(P,M\circ\Theta_E^n\circ \xi^{(n)})$$

Then, since the adversary can perform only individual measurements
with the subsequent classical processing, the conditions of
Theorem \ref{TheoCsiszar} are satisfied.\end{proof}

\section{On advantages of quantum key distribution over classical
one}

Thus, it is possible that for a single use of a quantum channel
the classical condition (\ref{EqCCCond}) of key distribution is
not satisfied, but with multiple repetitions Bob uses the
essentially quantum property, entangled measurements, in contrast
to the adversary Eve, who performs in essence classical attacks --
factorized measurements. As a result, Bob achieves advantage over
Eve and the subsequent key distribution according to the classical
scheme becomes possible.

For demonstration of this possibility we give an example of the
quantum key distribution system, in which at every run of the
quantum channel Eve receive the same state as Bob, i.e., there is
an ideal eavesdropping. However, as a consequence of the effect of
entangled measurements applying by Bob with the multiple use of
the channel, he gets more information than Eve, and the key
distribution becomes possible. The effect of entangled
measurements is caused by the use of nonorthogonal states.

\begin{example}Let $\mathcal H_A=\mathbb C^2\otimes\mathbb C^2$, $\mathcal
H_B=\mathcal H_E=\mathbb C^2$, $\Theta$ is an identical map.
$\mathcal A=\{0,1\}$,
$$\xi(0)=(|\varphi\rangle\langle\varphi|)^{\otimes 2},$$
$$\xi(1)=(|\psi\rangle\langle\psi|)^{\otimes 2},$$ where
$$|\varphi\rangle,|\psi\rangle\in\mathbb C^2,~~\langle\varphi|\psi\rangle\neq 0.$$
If for $P\in\mathcal{P(A)}$ the condition $P(0)=0$ or $P(1)=0$ is
satisfied, then $$\forall M_E\in\mathcal{M(H}_E;\mathcal E)$$ we
have
$$I(P,M_E\circ\Theta_E\circ\Xi)=0,$$ so $$\max_P
I(P,M_E\circ\Theta_E\circ\Xi)$$ is achieved when $P(0),P(1)>0$.
Then, due to the fact that the operators $P(0)\xi(0)$ and
$P(1)\xi(1)$ don't commute, condition (\ref{EqQCCond}) is
satisfied. So, by Theorem \ref{TheoOur} the key distribution is
possible.\end{example}

There aren't analogues in classical cryptography for the given
effect when the key distribution is possible even under conditions
of ideal eavesdropping. In this sense one can say about advantages
of quantum cryptography over classical one.

This work was partially supported by the Russian Foundation of
Basic Research (project 05-01-00884), the grant of the President
of the Russian Federation (project NSh-1542.2003.1) and the
program "Modern problems of theoretical mathematics" of the
Mathematical Sciences department of the Russian Academy of
Sciences.


\begin{thebibliography}{99}
\bibitem{BennettBrassard84}Bennett, C.H., Brassard, G. Quantum
cryptography: public key distribution and coin tossing.
Proceedings of IEEE International Conference on Computers, Systems
and Signal Processing, Bangalore, India, 1984, pp. 175 -- 179.
\bibitem{Mayers}Mayers, D. Unconditional security in quantum
cryptography. quant-ph/9802025
\bibitem{Biham}Biham, E., Boyer, M., Boykin, P.O., Mor, T.,
Roychowdhurry, V. A proof of the security of quantum key
distribution. quant-ph/9912053
\bibitem{LoChau}Lo, H.-K., Chau, H.F. Unconditional security of
quantum key distribution over arbitrarily long distances. Science,
1999, Vol. 283, pp. 2050 -- 2056. quant-ph/9803006
\bibitem{ShorPreskill}Shor, P.W., Preskill, J. Simple proof of security of the
BB84 quantum key distribution protocol. Physical Review Letters,
2000, Vol. 85, pp. 441 -- 444. quant-ph/0003004
\bibitem{Gisin}Gisin, N. et al. Quantum cryptography.
quant-ph/0101098
\bibitem{VolovichVolovich}Volovich, I.V., Volovich, Ya.I. On classical and quantum
cryptography. quant-ph/0108133
\bibitem{Zeilinger}Poppe, A. et al. Practical
quantum key distribution with polarization Entangled Photons.
quant-ph/0404115
\bibitem{BenorMayers}Ben-Or, M., Mayers, D. General security
definition and composability for quantum and classical protocols.
quant-ph/0409062
\bibitem{RennerEkert}Christandl, M., Renner, R., Ekert, A. A
generic security proof for quantum key distribution.
quant-ph/0402131
\bibitem{RennerGisin}Renner, R., Gisin, N., Kraus, B. An
information-theoretic security proof for QKD protocols.
quant-ph/0502064
\bibitem{OhyaPetz}Ohya M. Petz D. Quantum entropy and its use.
Berlin,
Springer-Verlag, 1993.
\bibitem{Holevo}Holevo A.S. Introduction to quantum information theory.
Moscow, MCNMO, 2002. (in Russian)
\bibitem{CsiszarKorner78}Csisz\'{a}r, I., K\"{o}rner, J. Broadcast channels
with confidential messages. IEEE Transactions on Information
Theory, 1978, Vol. 24, No 3, pp. 339 -- 348.
\end{thebibliography}
\end{document}